\documentclass[conference]{IEEEtran}
\IEEEoverridecommandlockouts
\usepackage{cite}
\usepackage{amsmath,amssymb,amsfonts}
\usepackage{algorithmic}
\usepackage{graphicx}
\usepackage{textcomp}
\usepackage{xcolor}
\usepackage{stfloats} 
\usepackage{tabularx}
\usepackage{subcaption}
\usepackage{hyperref}
\usepackage{fancyhdr}
\usepackage{booktabs}
\usepackage{multirow}
\usepackage{siunitx}
\usepackage{pdfpages}
\usepackage{url}
\usepackage{mwe, multicol}
\usepackage{float}
\usepackage{tikz}
\usepackage{textcomp}
\usepackage{lipsum}

\def\BibTeX{{\rm B\kern-.05em{\sc i\kern-.025em b}\kern-.08em
    T\kern-.1667em\lower.7ex\hbox{E}\kern-.125emX}}

\newcommand\copyrighttext{%
  \footnotesize \textcopyright 2023 IEEE. Personal use of this material is permitted.
  Permission from IEEE must be obtained for all other uses, in any current or future
  media, including reprinting/republishing this material for advertising or promotional
  purposes, creating new collective works, for resale or redistribution to servers or lists, or reuse of any copyrighted component of this work in other works.}
\newcommand\copyrightnotice{%
\begin{tikzpicture}[remember picture,overlay]
\node[anchor=south,yshift=10pt] at (current page.south) {\fbox{\parbox{\dimexpr\textwidth-\fboxsep-\fboxrule\relax}{\copyrighttext}}};
\end{tikzpicture}%
}

\makeatletter

\let\old@ps@IEEEtitlepagestyle\ps@IEEEtitlepagestyle
\def\confheader#1{%
    \def\ps@IEEEtitlepagestyle{%
        \old@ps@IEEEtitlepagestyle%
        \def\@oddhead{\strut\hfill#1\hfill\strut}%
        \def\@evenhead{\strut\hfill#1\hfill\strut}%
    }%
    \ps@headings%
}
\makeatother

\confheader{%
    \parbox{50cm}{\textbf{Accepted and presented at "2023 26th International Conference on\\ Computer and Information Technology (ICCIT) 13-15 December 2023, Cox’s Bazar, Bangladesh"}}
}

\begin{document}

\title{An Interpretable Deep Learning Approach for Skin Cancer Categorization\\

}

\author{\IEEEauthorblockN{Faysal Mahmud, Md. Mahin Mahfiz, Md. Zobayer Ibna Kabir, Yusha Abdullah}
\IEEEauthorblockA{\textit{Department of Computer Science and Engineering} \\
\textit{Ahsanullah University of Science \& Technology, Dhaka, Bangladesh}\\
\{faysalmmud, mmahinm10, ibnakabir081, yusha.abd\}@gmail.com
}
}

\maketitle
\copyrightnotice

\begin{abstract}
Skin cancer is a serious worldwide health issue, precise and early detection is essential for better patient outcomes and effective treatment. In this research, we use modern deep learning methods and explainable artificial intelligence (XAI) approaches to address the problem of skin cancer detection. To categorize skin lesions, we employ four cutting-edge pre-trained models: XceptionNet, EfficientNetV2S, InceptionResNetV2, and EfficientNetV2M. Image augmentation approaches are used to reduce class imbalance and improve the generalization capabilities of our models. Our models' decision-making process can be clarified because of the implementation of explainable artificial intelligence (XAI). In the medical field, interpretability is essential to establish credibility and make it easier to implement AI-driven diagnostic technologies into clinical workflows. We determined the XceptionNet architecture to be the best performing model, achieving an accuracy of 88.72\%. Our study shows how deep learning and explainable artificial intelligence (XAI) can improve skin cancer diagnosis, laying the groundwork for future developments in medical image analysis. These technologies' ability to allow for early and accurate detection could enhance patient care, lower healthcare costs, and raise the survival rates for those with skin cancer. Source Code: \href{https://github.com/Faysal-MD/An-Interpretable-Deep-Learning-Approach-for-Skin-Cancer-Categorization-IEEE2023.}{https://github.com/Faysal-MD/An-Interpretable-Deep-Learning-Approach-for-Skin-Cancer-Categorization-IEEE2023} 

\end{abstract}

\begin{IEEEkeywords}
Skin Cancer Detection, Deep Learning, Pre-trained Models, Convolutional Neural Networks (CNN), HAM10000, Medical Imaging, Explainable Artificial Intelligence (XAI)
\end{IEEEkeywords}

\section{INTRODUCTION}
Skin cancer may occur when skin cells are harmed, for as by excessive sunlight that produces ultraviolet (UV) radiation. \cite{pacheco2020impact}. Compared to other types of cancer, the majority of cancer statistics around the world do not require reporting of skin cancer diagnoses \cite{siegel2023cancer}. Nevertheless, according to the WHO, skin cancer accounts for one out of every three cancer diagnoses \cite{who}.
\par
There are many significant advantages to researching the elimination of skin cancer using computer vision and deep learning. These technologies enable earlier and more precise detection by automating the examination of enormous amounts of medical data, such as dermoscopic images. Particularly in areas with few medical resources, this can result in prompt interventions, better patient outcomes, lower healthcare costs, and increased access to professional diagnostics. Previous research overlooked the possibilities of explainable AI technology and instead concentrated only on computer vision and deep learning models. In contrast, our research applies explainable AI technology in addition to advanced computer vision and deep learning techniques for cancer image classification. With the help of this integration, we can classify cancer images accurately and achieve a thorough understanding of how the model makes decisions. By identifying the reasoning behind the model's conclusions, we increase the approach's interpretability, transparency, and reliability, which leads to more accurate and clinically useful outcomes in the field of medical image analysis. 
\par
This issue can be resolved by using an automated detection method based on machine learning. Deep learning is now frequently utilized for medical image processing because of its capacity to automatically learn complex characteristics from raw data. Numerous works, including the classification of cardiovascular disease, cancer, pneumonia \cite{10103293} and tumors \cite{shawon2023explainable}, have made significant use of deep learning in their works. In this article, we used the HAM10000 dataset to apply four pre-trained CNN models. The XceptionNet, EfficientNetV2S, EfficientNetV2M, and InceptionResNetV2 are well-liked deep learning models that are employed for this research. With an accuracy rate of 88.72\%, XceptionNet outperformed the other models. For categorizing images, numerous Explainable Artificial Intelligence systems have already been created. Identification and classification of skin cancer, however, have received less attention. Faster Score-CAM \cite{wang2020score} and SmoothGrad \cite{smilkov2017smoothgrad} were used in this article to correctly classify the images. 
\par
In this article, we present our contributions, which encompass the development of a skin cancer detection technique. To make up for the imbalance in the dataset, we utilized image augmentation. The pre-processed dataset was used to evaluate four pre-trained CNN models: XceptionNet, EfficientNetV2S, EfficientNetV2M, and InceptionResNetV2. Several performance measures, such as recall, precision, and the F1 Score of the weighted average, were used to examine the model's efficiency. Furthermore, to improve the interpretability and comprehension of the results, we used explainable artificial intelligence methods such as SmoothGrad and Faster Score-Cam to examine the predictions made by the trained models.

\section{LITERATURE REVIEW}
Innani et al. \cite{innani2022deep} came up with an idea to build an encoder-decoder architecture for the computer-aided diagnosis of skin cancers. A two-step framework is followed in this paper with the use of the HAM10000 dataset \cite{tschandl2018ham10000}. The first step is to segment the images and the second step is to remove redundant data and focus on the region of interest. The training is performed on deep learning models such as ResNet50, ResNet101, MobileNet, Xception, EfficientNetB0, and EfficientNetB3. Accuracy is used to evaluate classification performance, both with and without ROI. MobileNet has the highest accuracy of 78.54\% without ROI. On the other hand, Xception gives the top accuracy of 82.41\% with ROI. An artificial intelligence-based skin cancer detection system was created by Fraiwan et al. \cite{fraiwan2022automatic} using transfer learning with 13 deep convolutional neural network models.  All the models are trained with the HAM10000 \cite{tschandl2018ham10000} dataset. No extra feature extraction or segmentation is performed on this dataset. The DenseNet201 model achieves 73.5\% accuracy out of the 13 models implemented in this study. Nath et al. \cite{nath2023gastrointestinal} applied in their paper the pre-trained CNN for image classification task and Explainable AI (XAI) to ensure the model's reliability. The research by Kiran Pai and Anandi Giridharan \cite{pai2019convolutional} highlights that early detection and treatment of skin cancer offer a high possibility of survival. The researchers applied the VGGNet architecture as a Convolutional Neural Network (CNN) model for imbalanced multiclass classification of skin cancer. According to the research, after 50 epochs of training, the model obtained a test accuracy of 78\%. Huang et al. \cite{huang2021development} mentioned that skin cancer is the most prevalent type of cancer worldwide, not just in Taiwan. As a result, for the classification of skin lesions, they have developed a deep learning model that is both lightweight and efficient and that they can deploy on cloud platforms and mobile devices as well. For the classification, two datasets were employed. The HAM10000 dataset \cite{tschandl2018ham10000} is one of them. They evaluated the data using EfficientNet B1 to B7 models. They got the highest accuracy of 85.8\% from EfficientNet B4. Different Explainable AI strategies e.g. GRAD-CAM, Score-CAM have been used in both medical \cite{shawon2023explainable,van2022explainable} and non-medical works \cite{10103341} to interpret their models for different test samples.

\section{ BACKGROUND STUDY}
\subsection{XceptionNet}
XceptionNet \cite{chollet2017xception} is a convolutional neural network architecture with depth-wise separable convolutions that was designed for image classification applications. It was created by Google and differs from conventional CNNs in that separable convolutions according to depth, which divide the spatial and channel-wise operations, are used in place of standard convolutional layers. Reducing the number of parameters without sacrificing model performance is made possible by this separation, which improves computational efficiency. The architecture makes it possible for information to move through the network's layers more independently, which enhances learning capabilities as well as is centered on ensuring the effective use of depth-wise separable convolutions to capture intricate patterns. A sequence of convolutional building blocks, each consisting of depth-wise convolutions followed by point-wise convolutions, make up its architecture. XceptionNet can minimize computational overhead and learn hierarchical representations of input data. Even with its straightforward architecture, XceptionNet performs excellently on a range of image recognition benchmarks, proving how well it impacts a balance between classification accuracy and model efficiency.

\subsection{EfficientNetV2S}
EfficientNetV2S \cite{tan2021efficientnetv2}, is a deep learning architecture for image classification. It has a scaled version, signified as 'S,' that tries to achieve a balance between accuracy and model size. This version optimizes model performance by applying depth, width, and resolution scaling in a manner akin to the original EfficientNet's compound scaling technique. The 'S' variant maintains its efficacy in handling a variety of visual data while emphasizing computational efficiency by lowering the number of parameters. Utilizing a variety of cutting-edge methods, EfficientNetV2S prioritizes increasing training speed and memory efficiency. More complex modules, such as the Squeeze and Excitation blocks and the Swish activation function, are incorporated to improve feature representation and model expressiveness. In addition, it makes use of a redesigned architecture that prioritizes a multi-scale strategy and makes use of various receptive fields to efficiently capture both local and global features. EfficientNetV2S achieves competitive performance on multiple image classification benchmarks while preserving computational efficiency through an effective trade-off between model complexity and accuracy.

\subsection{InceptionResNetV2}
InceptionResNetV2 \cite{szegedy2017inception} is a mixture of the InceptionNet and ResNet designs, two well-known deep learning architectures. It has many convolutional layers, inception modules, and residual connections. The stem, inception blocks, reduction blocks, and fully connected layers make up the network architecture.
The first convolutional and pooling layers in the stem of InceptionResNetV2 are responsible for processing the input images. The fundamental building blocks, known as inception blocks, combine pooling operations with several parallel convolutional pathways with various filter sizes ($1\times1$, $3\times3$, and $5\times5$) to effectively extract a variety of characteristics. These building elements make feature extraction easier at various complexity levels and scales. Reduction blocks are also used to reduce the feature maps' spatial dimensions without losing any crucial information. To downsample the feature maps, these blocks combine convolutional layers with pooling processes. At the end of the network, fully connected layers are used to do classification using the features that have been extracted.

\subsection{EfficientNetV2M}
A version of the EfficientNet series, EfficientNetV2M \cite{tan2019efficientnet} aims to balance model size and accuracy in image classification applications. The 'M' stands for a scaled variant that uses compound scaling to improve performance by optimizing the depth, width, and resolution of the network. The goal of this model is to maximize efficiency by lowering computational requirements without sacrificing competitive accuracy. To improve its capabilities, it presents cutting-edge architectural innovations and creative training techniques. This architecture utilizes diverse techniques such as Swish activation functions and Squeeze-and-Excitation blocks, empowering the network to capture intricate features efficiently. EfficientNetV2M incorporates a redesigned network architecture with a focus on multi-scale feature fusion to efficiently handle local and global data. It demonstrates its skill at striking a balance between efficiency and effectiveness by achieving impressive accuracy benchmarks across a variety of image classification datasets by carefully managing the trade-offs between model complexity and performance.

\subsection{Explainable Artificial Intelligence}
Many Explainable Artificial Intelligence algorithms have already been developed for image categorization. Skin cancer identification and classification, however, have gotten less attention. Using SmoothGrad \cite{smilkov2017smoothgrad} and Faster Score-CAM \cite{wang2020score}, the interpretability of the proposed models was enhanced. In addition to reduce visual noise, SmoothGrad \cite{smilkov2017smoothgrad} is compatible with a variety of sensitivity map techniques. After calculating the gradient for several samples surrounding the given sample and incorporating data from Gaussian noise, the average is determined. A quicker version of Score-CAM is referred to as Faster Score-CAM \cite{wang2020score}. In comparison to Score-CAM, it is more effective and clarifies the model better. Faster Score-CAM employs the channels with substantial fluctuations as mask pictures since multiple picture channels greatly influence how the final heat map is created.

\subsection{Performance Metrics}
Different metrics, often known as performance metrics are used to evaluate the efficiency of the model. The performance of the model is evaluated using f1 score, recall, and precision \cite{vakili2020performance}. The formula we applied to calculate the f1 score, recall, and precision is given below:
    
    \begin{equation}
        Accuracy = \frac{(TP + TN)}{(TP + TN + FP + FN)}
    \end{equation}
    \begin{equation}
        Precision = \frac{TP}{(TP + FP)}
    \end{equation}
    \begin{equation}
        Recall = \frac{TP}{(TP + FN)}
    \end{equation}
    \begin{equation}
        F1 Score = 2 \times \frac{(Precision \times Recall)}{(Precision + Recall)}
    \end{equation}

\section{DATASET}
\begin{figure}[h]
\centerline{\includegraphics[width = 0.4\textwidth]{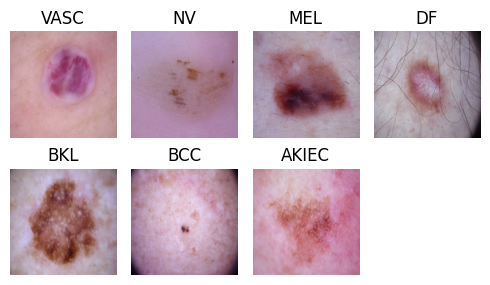}}
\caption{Visualization of skin cancer images from dataset}
\label{fig:cancerImages}
\end{figure}

The HAM10000 dataset (Human Against Machine) \cite{tschandl2018ham10000} was gathered from Harvard Dataverse. Melanoma (MEL), MelanocyticNevi (NV), BenignKeratosis (BKL), BasalCellCarcinoma (BCC), Vascular (VASC), ActinicKeratoses (AKIEC), and Dermatofibroma (DF) are the seven skin lesions contained in the collection, which has 10015 visuals of skin lesions with a resolution of $600\times450$ pixels. Figure \ref{fig:cancerImages} displays representations of 7 types of skin cancer collected from the HAM10000 database and the frequency is shown in Figure \ref{fig:ham} below:

\begin{figure}[h]
\centerline{\includegraphics[width = 0.4\textwidth]{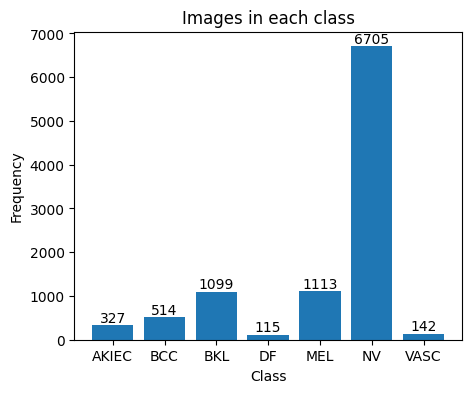}}
\caption{HAM10000 Data Frequency}
\label{fig:ham}
\end{figure}

\section{PROPOSED METHODOLOGY}
\begin{figure}[h]
    \centerline{\includegraphics[width = 0.47\textwidth]{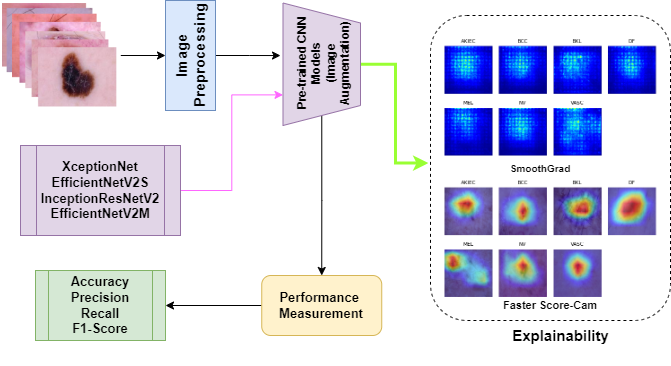}}
    \caption{Methodology}
    \label{fig:methodology}
\end{figure}

This section describes the proposed methodology for detecting skin cancer and the power of deep learning models. Data preprocessing, training process, and evaluation measures are all part of the methodology. This is a flow chart of the proposed methodology in Figure \ref{fig:methodology}. 

\subsection{Data Preprocessing}
We performed plenty of data preparation and augmentation in the first phase. The raw skin lesion images are downsized to $224\times224$ pixel resolution and labeled according to the skin cancer categories provided. The preprocessed images and labels are converted to a numpy array to improve data loading performance. The dataset is stratified and separated into 80:10:10 training, test, and validation sets. 

\begin{table}[!htbp]
    \caption{Class Wise Data Split Ratio}
    \label{tab:data-ratio}
    \centering
    \begin{tabular}{|l|c|c|c|c|}
        \hline
        \textbf{Class}  & \textbf{Train} & \textbf{Validation} & \textbf{Test} \\
        \hline
        AKIEC           & 264            & 30                  & 33  \\ \hline
        BCC             & 417            & 46                  & 51   \\ \hline
        BKL             & 890            & 99                  & 110    \\ \hline
        DF              & 93             & 10                  & 12    \\ \hline
        MEL             & 902            & 100                 & 111    \\ \hline
        NV              & 5430           & 604                 & 671    \\ \hline
        VASC            & 115            & 13                  & 14    \\ \hline
    \end{tabular}
\end{table}

Training data is augmented with various techniques such as rotations, flips, zooming, shearing, and shifting to improve model generalization. This phase ensures the availability of a well-structured, diverse, and balanced dataset for subsequent model training.

\subsection{Model Training and Optimization}
The four deep learning models are trained on the preprocessed and augmented training dataset during the second phase. Each model is set up with the Adam optimizer and a categorical cross-entropy loss function with a 0.001 learning rate. The goal of this phase is to train the models to learn discriminative features and patterns from skin lesion images, capturing the subtle specifics that distinguish different types of skin cancer. 

\begin{table}[!htbp]
    \caption{Models Configuration}
    \label{tab:Configuration}
    \centering
    \begin{tabular}{|l|c|c|c|c|}
        \hline
        \textbf{Model}     & \textbf{Dropout} & \textbf{Learning Rate} & \textbf{Epoch} & \textbf{Batch} \\
        \hline
        XceptionNet        & 0.5              & 0.001                  & 55             & 16 \\ \hline
        EfficientNetV2S    & 0.5              & 0.001                  & 50             & 16 \\ \hline
        InceptionResNetV2 & 0.5              & 0.001                  & 75             & 16 \\ \hline
        EfficientNetV2M    & 0.5              & 0.001                  & 80             & 16 \\ \hline
    \end{tabular}
\end{table}

A learning rate scheduler, particularly, the ReduceLROnPlateau callback is used in the training phase to achieve higher convergence and keep track of the validation accuracy. If the validation accuracy reaches a particular threshold (patience = 5), the learning rate is reduced. By carefully tweaking the learning rate throughout training, this strategy helps in refining the model's performance. A Model checkpoint callback is used to ensure that model progress is saved effectively. This callback tracks validation accuracy and stores the model with the best validation accuracy. This approach prevents any setbacks and allows for the retrieval of the optimal model configuration. By putting more layers on top of the base model, the architecture of the model is expanded. GlobalAveragePooling2D is applied to the output of the base model, and then a Dense layer with 128 units and ReLU activation is applied. A Dropout layer with a rate of 0.5 is implemented to prevent overfitting. At last, adding a Dense layer with a softmax activation function completes the model's output.

\subsection{Model Evaluation and Performance Comparison}
In the third phase, we evaluate each model's performance for the detection of skin cancer using important metrics such as accuracy, recall, precision, and F1-score. Based on its overall predicted performance and accuracy, this clear comparison assists in determining which model is the most successful.

\section{EXPERIMENTAL RESULT}
\begin{figure}[!htbp]
\centerline{\includegraphics[width = 0.45\textwidth]{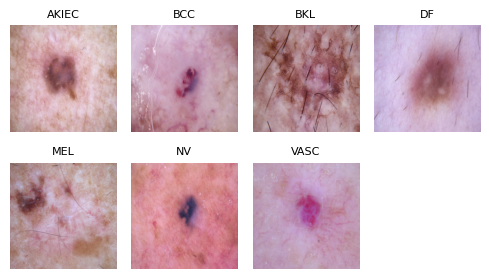}}
\caption{Classified skin cancer images from test data}
\label{fig:detected skin cancer}
\end{figure}

\begin{figure*}[htbp]
  \centering
  \subfloat[XceptionNet]{\includegraphics[width=1.7in]{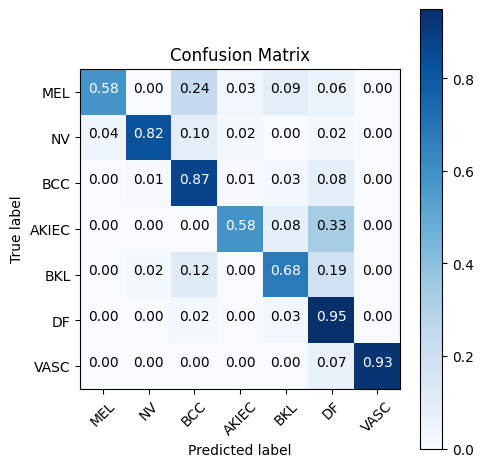}\label{fig:image1}}
  \hfil
  \subfloat[EfficientNetV2S]{\includegraphics[width=1.7in]{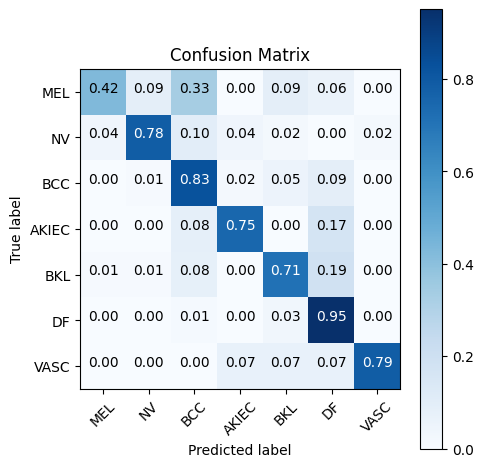}\label{fig:image2}}
  \hfil
  \subfloat[InceptionResNetV2]{\includegraphics[width=1.7in]{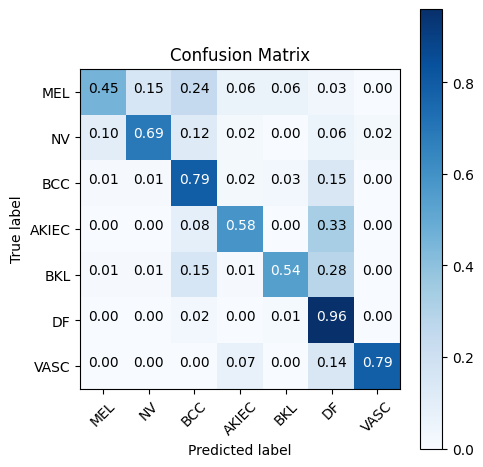}\label{fig:image3}}
  \hfil
  \subfloat[EfficientNetV2M]{\includegraphics[width=1.7in]{cm.png}\label{fig:image4}}
  
  \caption{Confusion Matrix of All Pre-trained Models}
  \label{fig:cm}
\end{figure*}

In this experiment, we use various pre-trained Convolutional Neural Network models for skin cancer detection such as XceptionNet, EfficientNetV2S, InceptionResNetV2, and EfficientNetV2M. Test accuracy, Explainability in Figure \ref{fig:c-ex-ai} and \ref{fig:mc-ex-ai}, confusion matrix in Figure \ref{fig:cm}, and a classification report with F1-score, precision, and recall for each class in Table \ref{tab:pm} are among the evaluation measures. The models detected 7 types of skin cancer in Figure \ref{fig:detected skin cancer}. 

\subsection{Result Analysis}
All four of the pre-trained models have been tested on a test set of seven different types of skin cancer images. All of these models predicted the cancer picture accurately. Table \ref{tab:pm} displays the findings from the test data using the four trained models.

\begin{table}[!htbp]
    \caption{Model-specific Classification Report of Weighted Average
}
    \label{tab:pm}
    \centering
    \begin{tabular}{|l|c|c|c|c|}
        \hline
        \textbf{Model}     & \textbf{Accuracy} & \textbf{Precision} & \textbf{Recall} & \textbf{F1-Score} \\
        \hline
        XceptionNet        & 88.72\%           & 0.89              & 0.89            & 0.89 \\ \hline
        EfficientNetV2S    & 88.02\%           & 0.88              & 0.88            & 0.88 \\ \hline
        InceptionResNetV2 & 85.73\%           & 0.86              & 0.86            & 0.85 \\ \hline
        EfficientNetV2M    & 85.02\%           & 0.89              & 0.89            & 0.89 \\ \hline
    \end{tabular}
\end{table}

\begin{figure*}[htbp]
  \centering
    \begin{subfigure}{\linewidth}
    \centering
    \includegraphics[width=3.6in]{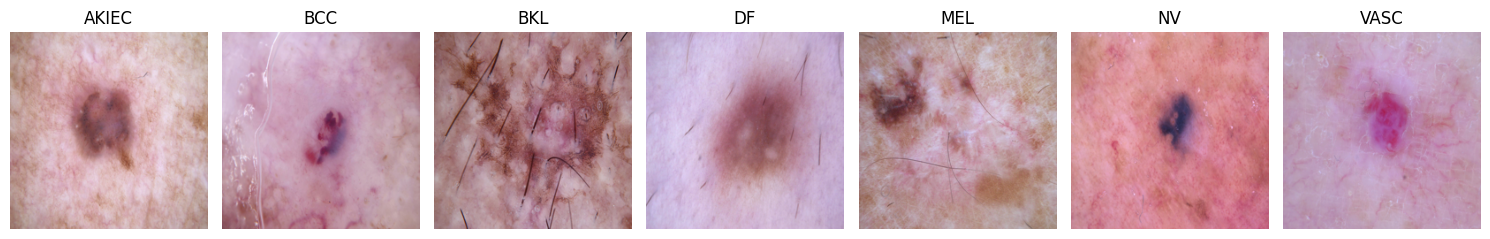}
    \caption{Original Image}\label{fig:env2s-ex-ai}
  \end{subfigure}
  
  \begin{subfigure}{\linewidth}
    \centering
    \includegraphics[width=3.6in]{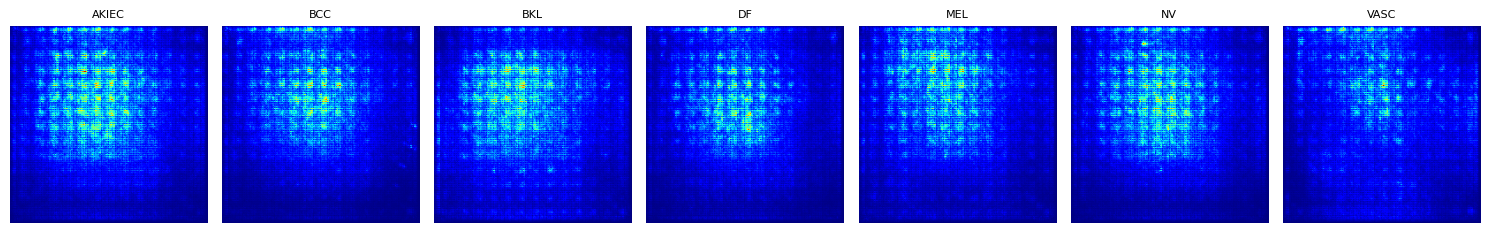}
    \caption{SmoothGrad Applied on XceptionNet}\label{fig:xn-ex-ai}
  \end{subfigure}
  
  \begin{subfigure}{\linewidth}
    \centering
    \includegraphics[width=3.6in]{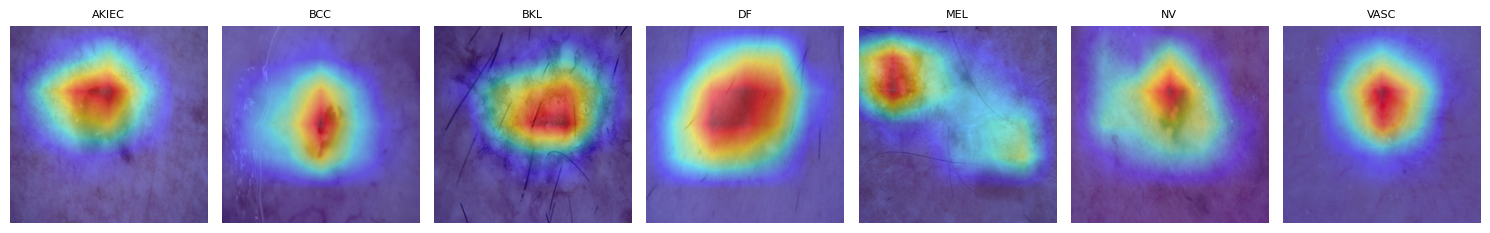}
    \caption{Faster Score-CAM Applied on EfficientNetV2S}\label{fig:}
  \end{subfigure}
  
  \begin{subfigure}{\linewidth}
    \centering
    \includegraphics[width=3.6in]{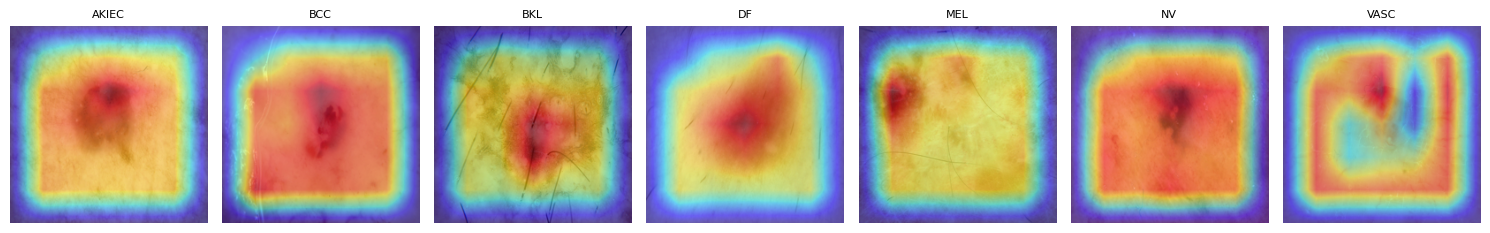}
    \caption{Faster Score-CAM Applied on EfficientNetV2M}\label{fig:env2m-ex-ai}
  \end{subfigure}
  
  \caption{Visualization of Classified Cancer Images using Explainable Artificial Intelligence (XAI)}\label{fig:c-ex-ai}
\end{figure*}

\begin{figure*}[htbp]
  \centering
  \begin{subfigure}{\linewidth}
    \centering
    \includegraphics[width=3.5in]{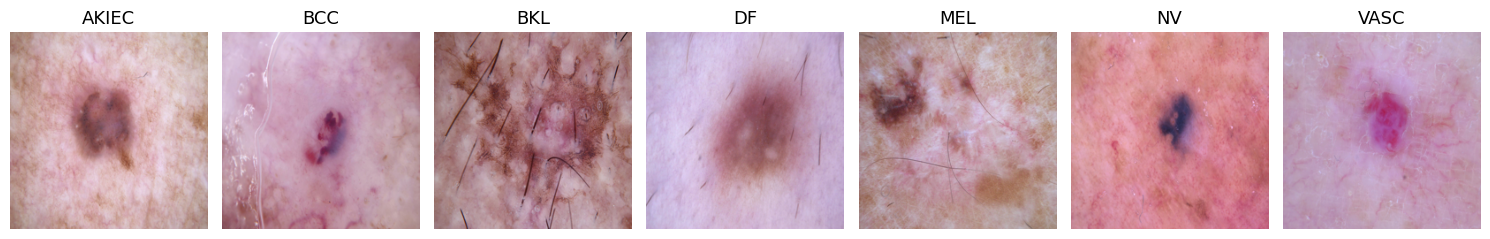}
    \caption{Original Image}\label{fig:env2s-ex-ai}
  \end{subfigure}
  
  \begin{subfigure}{\linewidth}
    \centering
    \includegraphics[width=3.5in]{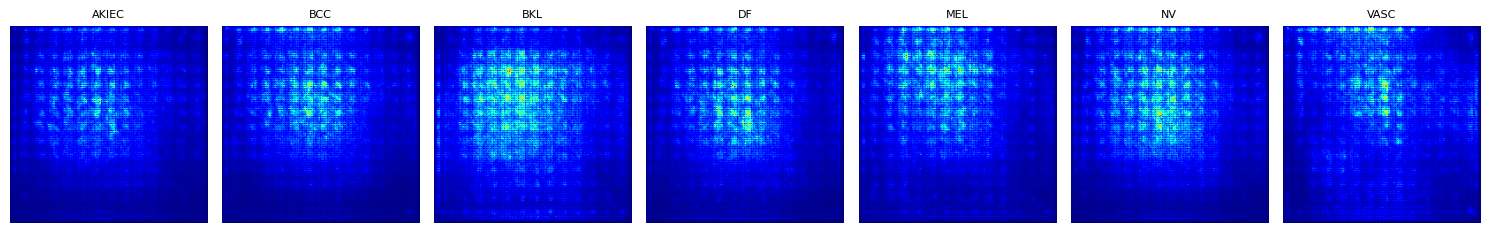}
    \caption{SmoothGrad Applied on XceptionNet}\label{fig:xn-ex-ai-mc}
  \end{subfigure}
  
  \begin{subfigure}{\linewidth}
    \centering
    \includegraphics[width=3.5in]{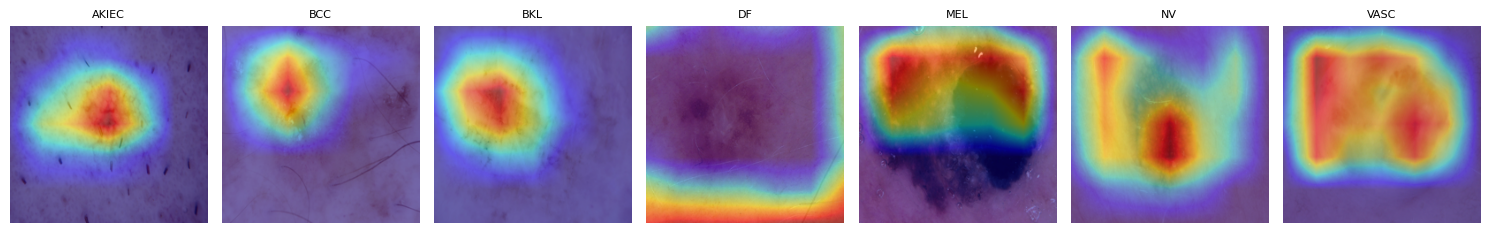}
    \caption{Faster Score-CAM Applied on EfficientNetV2S}\label{fig:irnv2s-ex-ai-mc-1}
  \end{subfigure}
  
  \begin{subfigure}{\linewidth}
    \centering
    \includegraphics[width=3.5in]{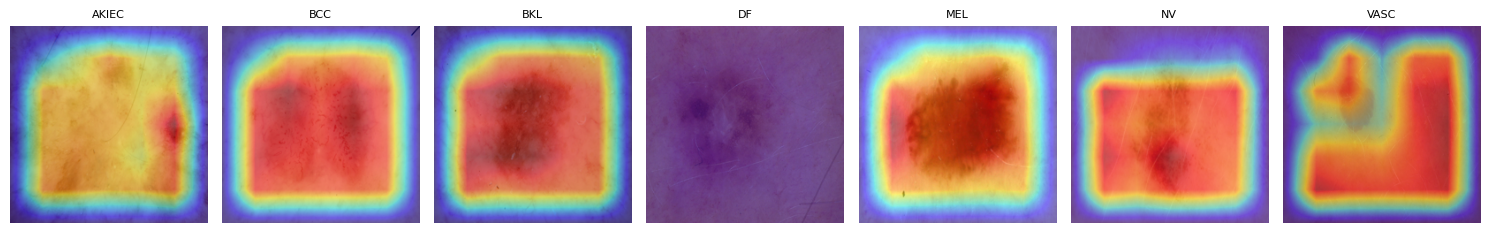}
    \caption{Faster Score-CAM Applied on EfficientNetV2M}\label{fig:irnv2s-ex-ai-mc-2}
  \end{subfigure}
  
  \caption{Visualization of Misclassified Cancer Images using Explainable Artificial Intelligence (XAI)}\label{fig:mc-ex-ai}
\end{figure*}

For all the models, we have used explainable artificial intelligence in Figure \ref{fig:c-ex-ai} and Figure \ref{fig:mc-ex-ai} to highlight the specific regions of cancer cells which can help to get a better image analysis. EfficientV2S and EfficientV2M are effective at classifying skin cancer images but occasionally are unable to correctly classify DF and MEL classes. In contrast, XceptionNet uses SmoothGrad to identify important regions of interest, but its ability to explain the behavior of the model is still relatively constrained. However, it falls short of providing a thorough understanding of the complex factors influencing classification outcomes. This leaves room for further investigation and improvement of interpretability techniques for improving model transparency and performance evaluation.

\section{RESULT COMPARISON}
In this section, we analyzed our obtained results and compared them to earlier research as well as the results obtained by deep learning models. The accuracy of earlier efforts that utilized deep learning models including MobileNet, Xception, EfficientNet, DenseNet, and VGGNet is displayed in Table \ref{tab:comparison}. Table \ref{tab:pm} shows the outcomes of our suggested model utilizing XceptionNet, EfficientNetV2S, InceptionResNetV2, and EfficientNetV2M. 

\begin{table}[!htbp]
    \caption{RESULT COMPARISON WITH SOME EARLIER PAPER}
    \label{tab:comparison}
    \centering
    \begin{tabular}{|l|c|c|c|}
        \hline
        \textbf{Paper}                  & \textbf{Method}        & \textbf{Accuracy} \\
        \hline
        \cite{innani2022deep}           & MobileNet              & 78.54\%  \\ \hline
        \cite{fraiwan2022automatic}     & DenseNet201            & 73.5\%  \\ \hline
        \cite{pai2019convolutional}     & VGGNet                 & 78\%  \\ \hline
        \cite{huang2021development}     & EfficientNetB4         & 85.8\%  \\ \hline
        \textbf{Proposed Model}         & \textbf{XceptionNet}   & \textbf{88.72\%}  \\ \hline
    \end{tabular}
\end{table}

The effectiveness of several neural network architectures was applied to a particular dataset in the comparative study of various approaches for cancer categorization that are currently in use. DenseNet201 and VGGNet had accuracies of 73.5\% and 78\%, respectively, while MobileNet attained an accuracy of 78.54\%. With an accuracy of 85.8\%, the EfficientNetB4 performed better than these models. Notably, with an outstanding accuracy of 88.72\%, the proposed XceptionNet model outperformed the other models. Given the dataset, this indicates that XceptionNet could be a viable option for cancer classification tasks, outperforming other well-known architectures such as MobileNet, DenseNet201, VGGNet, and EfficientNetB4.

\section{CONCLUSION}
Our goal in this article was to determine multiclass cancer lesions using the proposed methodology. We have used augmentation to resolve the imbalance problem in the HAM10000 dataset \cite{tschandl2018ham10000}. Then XceptionNet, EfficientNetV2S, EfficientNetV2M, and InceptionResNetV2 models were trained using the augmented dataset. After training the models, we achieved 85.03\% on EfficientResNetV2M, 85.73\% on InceptionResNetV2, 88.02\% on EfficientNetV2S, and finally, we gained an accuracy of 88.72\% for XceptionNet model. To gain a deeper understanding of the trained models, confusion matrices were also used which evaluated the performance of the models. By analyzing the accuracy and the performance metrics, we can say that computer vision can be a reliable method to detect skin malignancies instead of relying on inaccurate visual dermoscopic diagnosis. Our ultimate objective is to improve the accuracy. So that we can get a further accurate prediction from the models and make it a more reliable source. In the future, we would like to work with more additional samples of malignant skin lesions to the dataset which may give us a better accuracy. Therefore, early skin cancer identification can help a patient get better medical care and even save more lives.

\bibliographystyle{IEEEtran}
\bibliography{references}

\vspace{12pt}
\color{red}

\end{document}